\documentclass[preprint,12pt]{elsarticle}

\usepackage{graphicx}
\usepackage{dcolumn}
\usepackage{bm}
\usepackage{subcaption}
\usepackage{hyperref}
\usepackage{amsmath, amssymb}

\begin{document}

\title{End-to-End Jet Classification of Quarks and Gluons with the CMS Open Data
}

\author{M. Andrews$^1$, J. Alison$^1$, S. An$^{1,2}$, B. Burkle$^3$, S. Gleyzer$^4$, M.~Narain$^3$, M. Paulini$^1$, B. Poczos$^5$, E. Usai$^3$}
\address{$^1$ Department of Physics, Carnegie Mellon University, Pittsburgh, USA}
\address{$^2$ CERN, Geneva, Switzerland}
\address{$^3$ Department of Physics, Brown University, Providence, USA}
\address{$^4$ Department of Physics and Astronomy, University of Alabama, Tuscaloosa, USA}
\address{$^5$ Machine Learning Department, Carnegie Mellon University, Pittsburgh, USA}

\date{\today}

\begin{abstract}
We describe the construction of novel end-to-end jet image classifiers to discriminate quark- versus gluon-initiated jets using the simulated CMS Open Data. These multi-detector images correspond to true maps of the low-level energy deposits in the detector, giving the classifiers direct access to the maximum recorded event information about the jet, differing fundamentally from conventional jet images constructed from reconstructed particle-level information. Using this approach, we achieve classification performance competitive with current state-of-the-art jet classifiers that are dominated by particle-based algorithms. We find the performance to be driven by the availability of precise spatial information, highlighting the importance of high-fidelity detector images. We then illustrate how end-to-end jet classification techniques can be incorporated into event classification workflows using Quantum Chromodynamics di-quark versus di-gluon events. We conclude with the end-to-end event classification of full detector images, which we find to be robust against the effects of underlying event and pileup outside the jet regions-of-interest.
\end{abstract}

\maketitle


\section{\label{sec:Introduction}Introduction}
The study of jet substructure at the CERN Large Hadron Collider (LHC) has played an instrumental role in the understanding of the standard model (SM) of particle physics through the analysis of jets produced from Quantum Chromodynamics (QCD)~\cite{jetshapeatlas,jetshapecms} and from the decay of boosted heavy resonances or particles such as the top quark~\cite{ttsubstruct_atlas,ttsubstruct_cms,toptagatlas,deepcsv} or the Higgs boson~\cite{boostedhatlas,boostedhcms}. Furthermore, jet substructure analysis remains a central tool in searches for physics beyond the standard model (BSM) involving boosted heavy new resonances (see~\cite{jetsubstructrev,boostedrev} for comprehensive reviews). A major component of understanding jets arising from boosted heavy resonances, exotic or other BSM physics processes, is a detailed understanding of the vastly more dominant QCD jet background. For this reason, the characterization and discrimination of light quark- versus gluon-initiated jets that comprise QCD jets has also been extensively examined~\cite{atlasjetimg,deepjet}. Indeed, a number of BSM resonances involve preferential decays to either quarks or gluons~\cite{qgshapeslhc}, making their study valuable in their own right. The development of effective tools to identify gluons from their light-quark counterparts---and to analyze jet substructure in general---has become a popular topic of study as a result.

Differences in quark versus gluon jet radiation patterns arise from their distinct QCD color charges: gluon initiated jets carry a larger QCD color factor leading to a shower evolution with a higher branching probability~\cite{pdgreview,qgtheory}. The gluon jet thus exhibits a slightly softer and broader radiation pattern of higher particle multiplicity relative to a quark jet. To better appreciate the recent experimental advances in jet substructure discrimination that exploit this signature, it is instructive to sketch the jet finding and reconstruction process at the typical LHC experiment. Fundamentally, data recorded with the LHC experiments exists as detector-level information, either as energy depositions in calorimeter cells or ``hits'' in tracking systems such as pixel or strip detectors. Through a largely rule-based ``particle-flow'' algorithm~\cite{atlaspf,pf}, the detector data from the various subdetectors are combined and processed to reconstruct the particle-level information corresponding to the 4-momenta of all the resolved particles originating from a proton-proton collision at the LHC. A jet clustering algorithm~\cite{antikt} is then applied to group nearby particles into a jet object. A number of jet-level features can be derived to characterize the jet as a whole, such as its shape and particle multiplicity. At each stage of the jet reconstruction chain, a level of abstraction occurs to idealize the representation of the data. While this simplifies their interpretation, it potentially involves information loss that may be of disadvantage for more exhaustive BSM searches, where exotic decay signatures might be expected. 

While earlier techniques in jet substructure analysis involving LHC-like data began with jet-level variables~\cite{jetshapeatlas,jetshapecms,jetsubopendata}, recent advances in machine learning (ML)---particularly in deep neural networks (NN)---have permitted classification algorithms to directly probe the particle-level data constituting the jet. For a review of current state-of-the-art ML-based jet classification see~\cite{jetmlrev,toptagrev}. A number of non-trivial challenges have arisen from such approaches, for example, the order in which the particle-level data should be presented to the classifier, and how variable numbers of particles are handled. Especially for heavy resonance decays, classifier performance may be particularly sensitive to the choice of the ordering scheme. To address these issues, solutions have been proposed involving artificial but physically-motivated ordering schemes~\cite{deepcsv,atlasbjetrnn,jetflavorFCN,toptagFCN,toptagRNN,kyle,taoli,lolaqg}, as well as those utilizing order-invariant algorithms~\cite{eflow,particlecloud}. Another highly popular approach has been to exploit the spatial distribution of the particles in the detector as a natural solution to the ordering problem. Inspired by the immense success of convolutional neural networks (CNN) in computer vision~\cite{alexnet}, such ``jet images'' have been created by pixelating the particle-level data into a grid that amounts to a coarse-grained histogram-like image of the underlying detector geometry, which can then be fed to a CNN~\cite{atlasjetimg,jetimages,jetimages2,qvgimg}. However, as the pixelation process itself involves an additional, lossy operation, such attempts have been noted to underperform relative to algorithms that directly use particle-level data, even when artificial ordering is involved~\cite{deepjet, kyle}.

However, the ultimate information bottleneck in all of the above particle-data based approaches is the rule-based particle reconstruction algorithm itself. Most of the breakthrough success in applying CNNs to computer vision have come from bypassing rule-based ``feature engineering'' altogether and instead allowing CNNs to learn relevant features directly from the raw camera data. Therefore, this fact motivates the idea of applying ``end-to-end physics classification'' whereby CNNs are used to directly train on maps of the true detector-level data (or their simulated equivalents), in all their richness, \emph{before} any particle processing is performed. In theory, this gives the classifier full access to the maximum recorded event information at a level not achievable with processed particle- or jet-level data, while avoiding the particle ordering problem altogether. Thus, while the end-to-end approach bears a resemblance to existing jet image techniques, the underlying information content is fundamentally different. Among simpler neutrino experiments, this end-to-end strategy has quickly found success~\cite{neutrino,microboone}, but among the more mature LHC experiments, adoption is still in its infancy. Its success is by no means guaranteed, due to the much more complex detector systems and workflows of hadron collider experiments.

In previous work~\cite{acat,e2e}, we laid the basic foundation for performing end-to-end physics classification on complex LHC detectors like the Compact Muon Solenoid (CMS) using electromagnetic objects. In this paper, we apply the same techniques to the problem of jet substructure classification, again using simulated CMS Open Data taking advantage of this excellent public resource, as others have~\cite{jetsubopendata,circles}. We study the classification of jets initiated by quarks versus gluons from QCD dijet production and 
use image windows around the jet region-of-interest (ROI), or jet-view images, as done by the cited work above. While the image representation of the tracking information in Ref.~\cite{e2e} is simplistic, for our purposes, these images are not expected to significantly impact the classification of quark versus gluon jets that generally do not contain displaced secondary track vertices. We thus benchmark our results against one of the current state-of-the-art jet classifiers based on particle-level data~\cite{kyle}.

In addition, we consider the event classification of the full dijet process as either di-quark or di-gluon initiated. A number of potential implementations for this approach are explored: from more standard workflows, where the jet classification is factorized from the event classification, to a unified event classification workflow, where a single training is performed on the full detector-view images, analogous to~\cite{e2e} (and to some degree, also~\cite{circles,wahid,atlaspuevent}). While dijet decays are simple enough that jet ordering becomes trivial, they are useful as a pedagogical example to highlight the ability of end-to-end event classifiers on full detector-view images to completely capture discriminating features at both the jet- and event-level scales in the limit that the event topology can be fully and intuitively modeled by hand. A more general application demonstrating the performance of this technique for the case of complex multi-body decays in which both ordering and modeling are non-trivial is left for future work.

This paper is arranged as follows. In Section~\ref{sec:Samples}, we introduce our data sample and event selection, while in Section~\ref{sec:Images}, we briefly describe the CMS geometry and our jet image construction procedure. In Section~\ref{sec:Net}, we outline our network architecture and training strategy. Finally, in Sections~\ref{sec:jetID} and \ref{sec:eventID}, we present the results of our jet and event classification studies, respectively, and summarize our conclusions in Section~\ref{sec:Conclusions}.

\section{\label{sec:Samples}Open Data Simulated Samples}
We use simulated 2012 CMS Open Data~\cite{opendata} for all the studies presented in this paper. The CMS Open Data are ideally suited to the end-to-end approach due to their use of the \texttt{Geant4}~\cite{geant4} package, which delivers the state-of-the-art in first-principles detector simulation, together with the most detailed geometry models of the CMS detector available~\cite{fullsim}.

Both quark and gluon samples are derived from the QCD dijet production dataset with a generated invariant transverse momentum in the range $\hat{p}_T = 90 - 170\,\mathrm{GeV}$~\cite{dijet}. The events are generated and hadronized with the \texttt{Pythia\,6}~\cite{pythia6} package with the Z2* tune, which accounts for the differences in the quark versus gluon shower evolution. The samples also account for the multi-parton interactions from the underlying event and have run-dependent pileup (PU) ranging from a peak average PU of $\langle \mathrm{PU} \rangle = 18-21$.

We impose a basic event selection on the dijet events for them to be used and categorized under the quark or gluon class label, for both jet and event classification. First, we require that the two outgoing partons from the \texttt{Pythia} hard-scatter event be either both light quarks $q_l$, where $l=u,d,s$ (or their antiparticles), or both gluons, otherwise the event is discarded. In addition, each of the partons must be matched to a reconstructed jet to within a cone of $\Delta R<0.4$, where $R$ is the angular separation in the pseudorapidity-azimuthal ($\eta-\phi$) plane. Jets with wide-angle radiation are thus excluded from this study, for simplicity. In turn, each of the parton-matched reconstructed jets must have a minimum transverse momentum of $p_T > 70\,\mathrm{GeV}$ and pseudorapidity of $|\eta| < 1.8$. For the jet identification studies, only the leading-$p_T$ jet in each passing event is used.

For computational convenience, we only use a subset of the total dijet dataset of the CMS Open Data---we found no statistically significant advantage to training on the full dataset. In addition, to minimize training bias due to unbalanced class or PU representation, we truncate the dataset to contain a balanced proportion of samples per class per PU run era (Run2012AB, Run2012C, Run2012D), as summarized in Table \ref{table:Nevents}. We therefore obtain a grand total of 793900 samples for training and validation, and 139306 samples for the final test set. This thus gives our results in Sections~\ref{sec:jetID} and \ref{sec:eventID} a statistical uncertainty of about $0.4\%$, assuming Poisson statistics.

\begin{table}[!htbp]
\centering
\begin{tabular}{l c c c}
\hline
\textbf{Category} & \textbf{Run2012AB} & \textbf{Run2012C} & \textbf{Run2012D} \\
\hline
Train+Validation & 107778 & 148990 & 140182 \\
Test             & 18136  & 23770  & 27747 \\
\hline
\end{tabular}
\caption{Number of training events per class passing event selection, by PU run era. Only the leading-$p_T$ jet in the event is used for the jet classification studies.}
\label{table:Nevents}
\end{table}

\section{\label{sec:Images} CMS Detector and Images}

The CMS detector is arranged as a series of concentric cylindrical sections split into a barrel section and two circular endcap sections. The innermost sections comprise the inner tracking system for identifying charged particle tracks. This is then enclosed by the electromagnetic calorimeter (ECAL) which measures energy deposits from electromagnetic particles, followed by the hadronic calorimeter (HCAL) which measures energy deposits from hadrons. Finally, the calorimeters are enclosed by the outer tracking system used to identify muons.

The CMS Open Data contains the calibrated, reconstructed hits~\cite{cms} of the ECAL and HCAL at the crystal- and tower-level, respectively. Following the image construction techniques in~\cite{e2e}, we are able to form calorimeter images whose pixels correspond exactly to physical crystals or towers. Due to the complexity of the CMS tracking detectors and the absence of a
reconstructed hit collection for tracks in the CMS Open Data, the track information can only be approximated with the results of the
reconstructed track fits that are the only track collection present in the CMS Open Data. An ECAL-like granularity image grid is thus filled with the ($\eta$,$\phi$)-positions of each reconstructed track with pixel intensity equal to the track's $p_T$. A study into more accurate representations of the tracking information is reserved for future work. In practice, as long as the track image resolution is comparable to the effective resolution of the reconstructed track position, any loss in information from using a discrete image grid for the track positions is expected to be negligible. The full detector-view images therefore consist of three subdetector channels: one each for ECAL, HCAL, and the reconstructed tracks.

\begin{figure}[!htbp]
\centering
\begin{subfigure}{.78\textwidth}
  \centering
  \includegraphics[width=\textwidth]{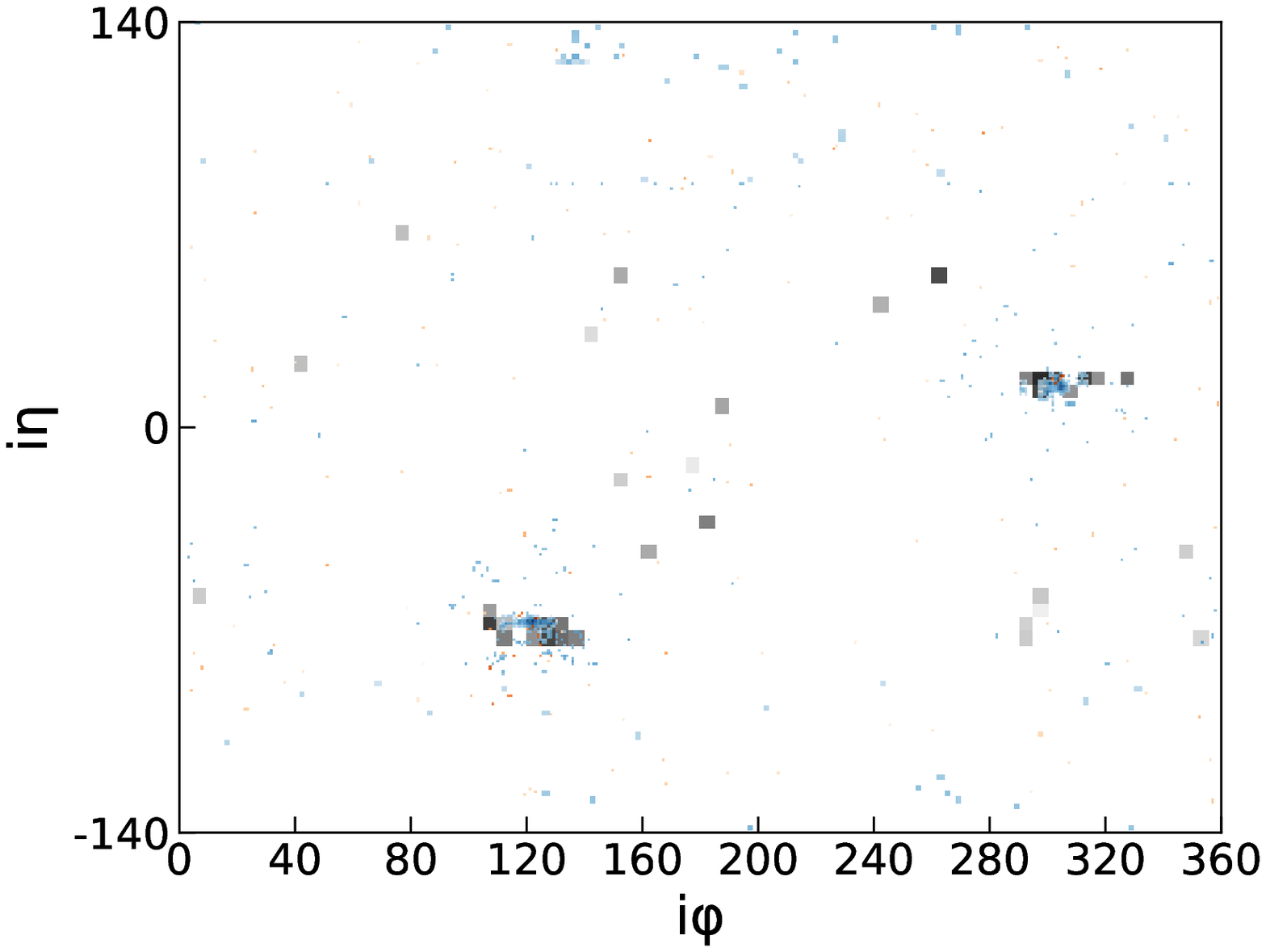}
  \caption{Composite full detector-view image. Image resolution: 280 $\times$ 360.}
  \label{fig:fullye2e}
\end{subfigure}
\begin{subfigure}{.9\textwidth}
  \centering
  \includegraphics[width=.45\linewidth]{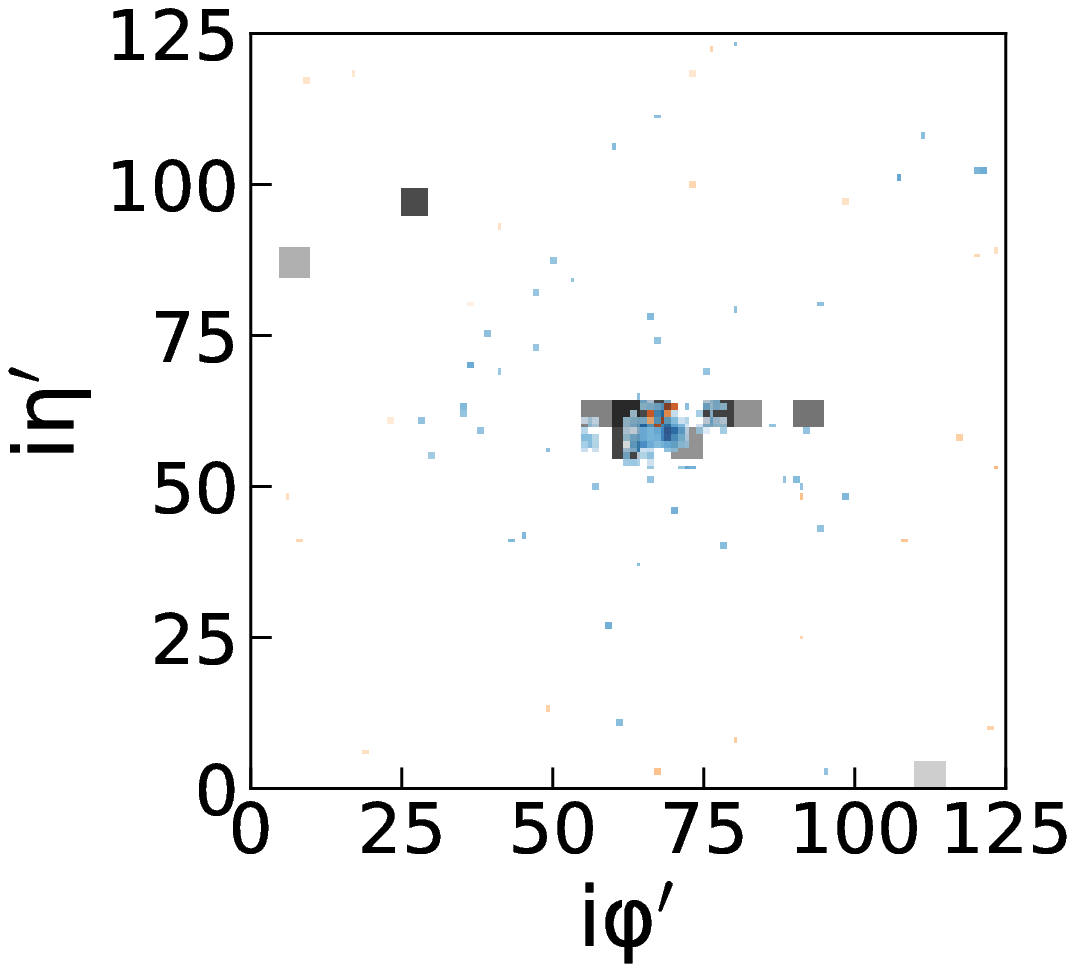}
  \includegraphics[width=.45\linewidth]{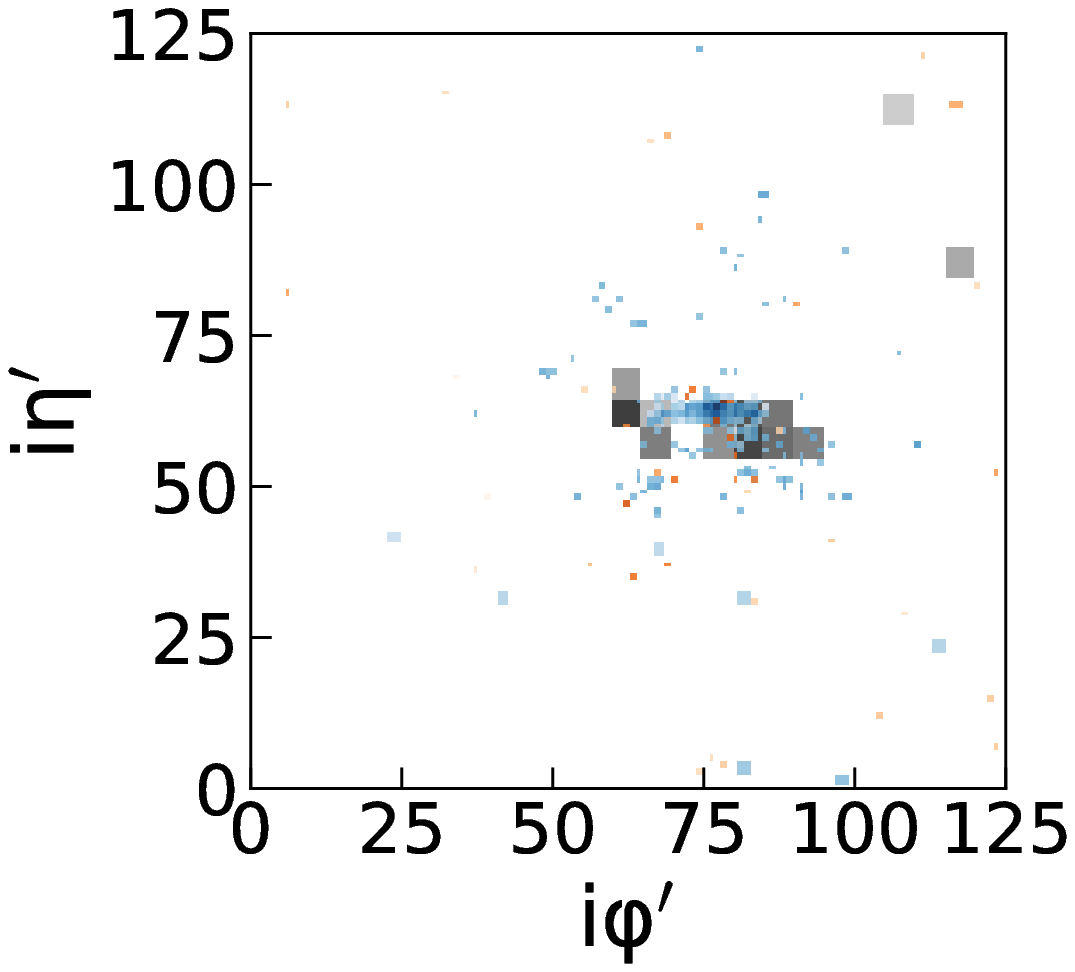}
  \caption{Composite individual jet-view images for leading jet (left) and sub-leading jet (right) in a representative dijet event. Image resolution: 125 $\times$ 125.}
  \label{fig:jetcrops}
\end{subfigure}
\caption{Representative dijet event in HCAL-centric geometry at EB-like granularity: full detector-view image (\ref{fig:fullye2e}) spanning the range $|\eta|<3$, and individual jet-view images (\ref{fig:jetcrops}) spanning $\Delta R \sim 1$. Images are multi-channel composites of information from tracks (orange), ECAL (blue), and HCAL (gray), all in log-scale.}
\label{fig:evt_nets}
\end{figure}

More specifically, for the barrel sections, the images are resolved in the finer ECAL barrel (EB), with the HCAL barrel (HB) hits up-sampled to match. The reconstructed track hits span the width of a single EB crystal. The difference in segmentation between the ECAL endcaps (EE) ($iX,iY$) and the HCAL endcaps (HE) ($i\eta,i\phi$) imposes a constraint on the construction of high-fidelity, multi-channel detector images. As explained in~\cite{e2e}, we devise two image geometry strategies: one where the EE segmentation is preserved and the HE hits are projected onto an ($iX,iY$) grid (ECAL-centric), and another where the HE segmentation is preserved and the ECAL hits are projected onto an ($i\eta,i\phi$) grid but at a finer EB-like granularity (HCAL-centric). In either case, the track hits appear as isolated pixels in the corresponding segmentation. As illustrated in Figure~\ref{fig:fullye2e}, this gives us a contiguous detector image of resolution $\Delta i\eta \times \Delta i\phi = 280 \times 360$ of ECAL barrel-like granularity, spanning the pseudorapidity range $|\eta| < 3$. For the entirety of this paper, we use only the HCAL-centric geometry strategy to simplify the construction of jet image windows. We found no significant benefit to appending additional wrap-around pixels along the $\phi$ edges of these full detector-view images. For the jet-view images, this is described below. 

The process of creating a jet-view image window given a reconstructed jet passing the event selection is as follows: from the full multi-subdetector image described above, we localize the jet by first taking the centroid of the reconstructed jet, then scan for the HCAL tower with the highest energy deposit within a window of $9\times9$ HCAL towers ($45\times45$ image pixels or $\Delta R \lesssim 0.4$). The most energetic HCAL tower defines the center of the jet-view image around which we crop a window of $125 \times 125$ image pixels ($\Delta R \lesssim 1$), as illustrated in Figure~\ref{fig:jetcrops}. For jets which fall near the $\phi$ edge of the detector image, we do a wrap-around padding so that the jet shower appear seamless and the jet-view image dimensions are preserved. We do not pad in the $\eta$ direction although this is, of course, a possibility. This imposes an effective pseudorapidity cut on the reconstructed jet of about $|\eta| < 1.57$. While the jet-view window sizes here were chosen to be fairly generous, future applications may choose to optimize the window sizes further to extend the jet pseudorapidity range.

\begin{figure}[!htbp]
\centering
\begin{subfigure}{.8\textwidth}
  \centering
  \includegraphics[width=.45\linewidth]{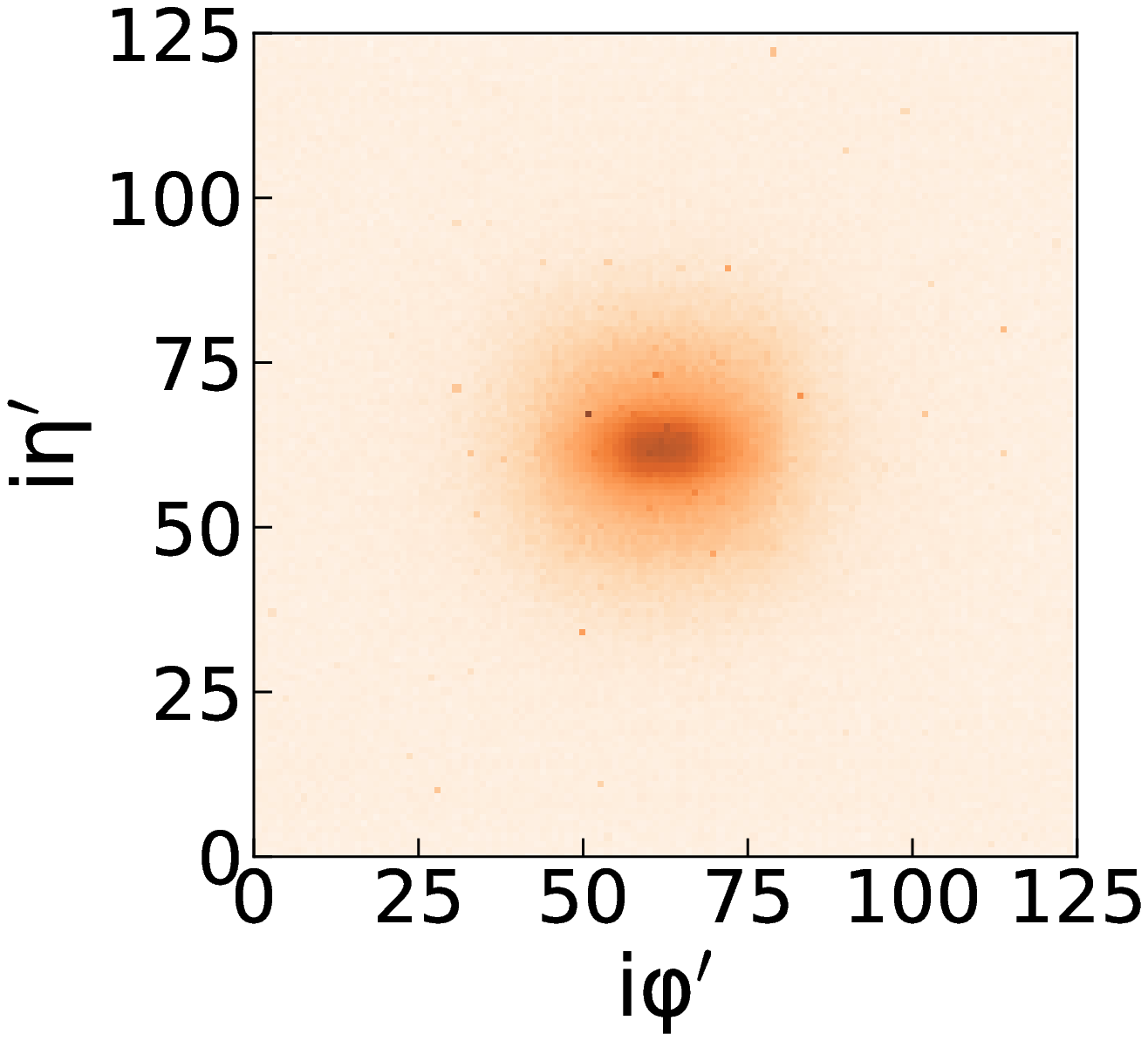}
  \includegraphics[width=.45\linewidth]{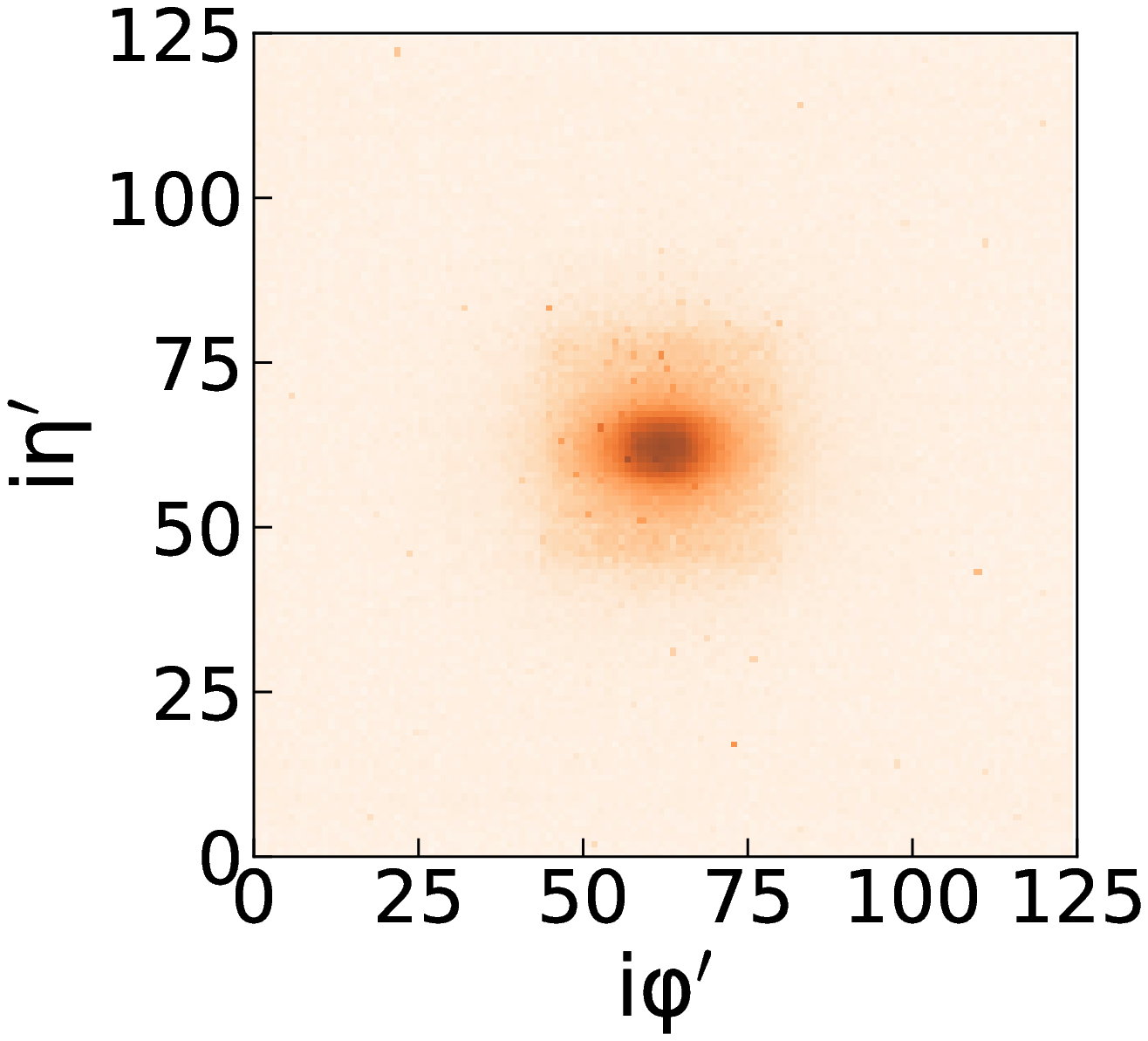}
  \caption{Tracks overlay. Left: gluon jet, Right: quark jet.}
  \label{fig:overlay_0}
\end{subfigure}
\begin{subfigure}{.8\textwidth}
  \centering
  \includegraphics[width=.45\linewidth]{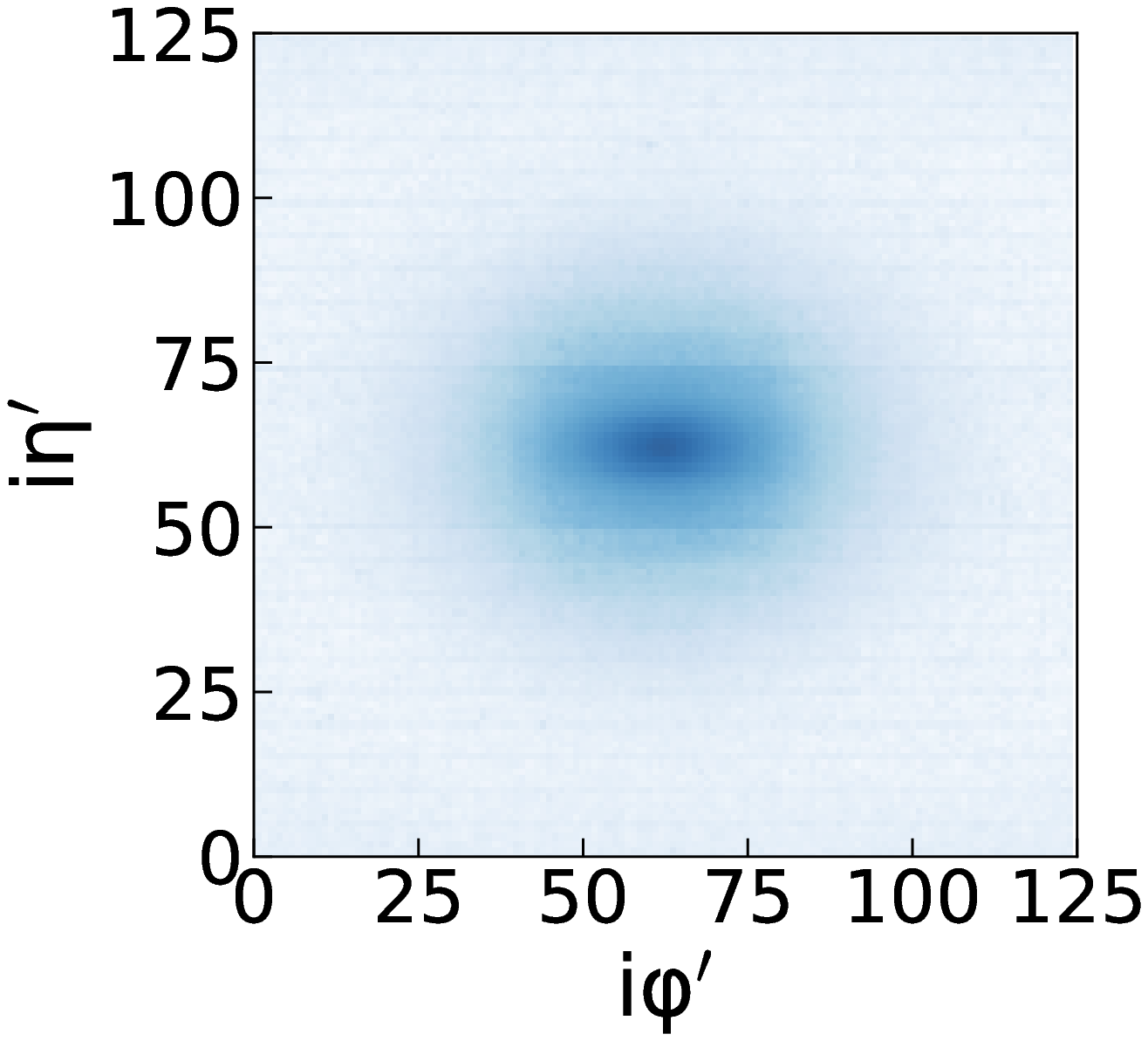}
  \includegraphics[width=.45\linewidth]{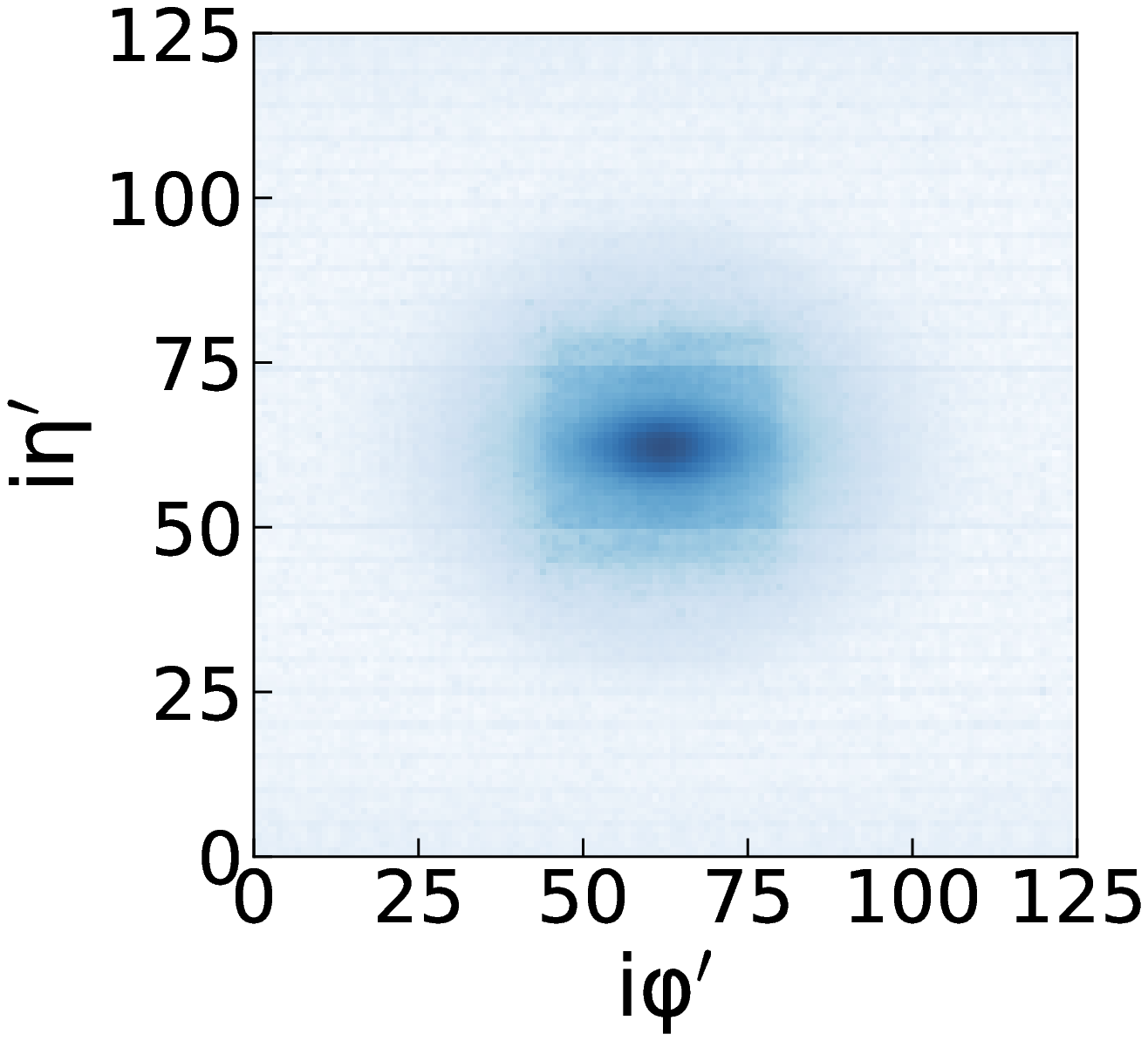}
  \caption{ECAL overlay. Left: gluon jet, Right: quark jet.}
  \label{fig:overlay_1}
\end{subfigure}
\begin{subfigure}{.8\textwidth}
  \centering
  \includegraphics[width=.45\linewidth]{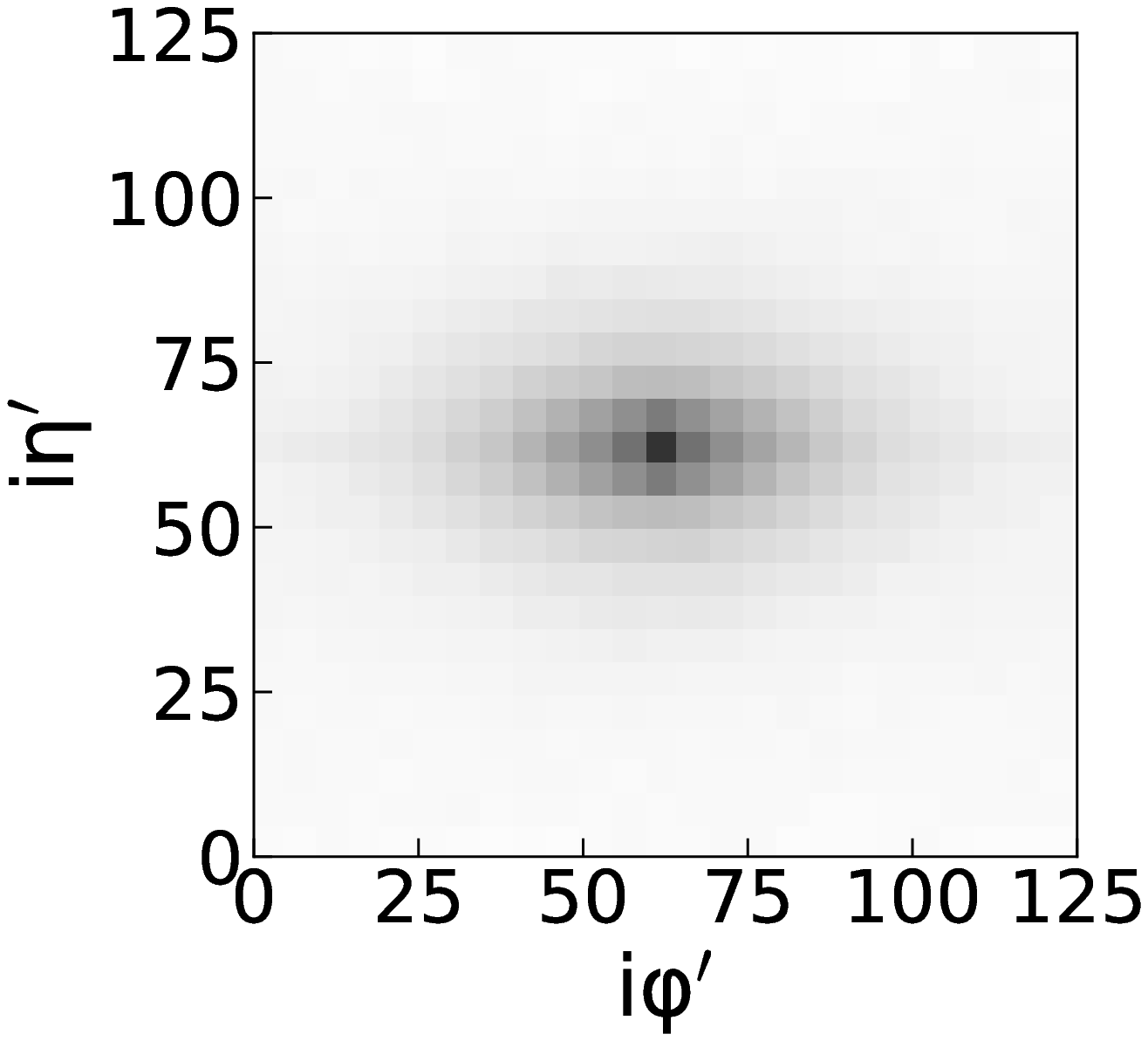}
  \includegraphics[width=.45\linewidth]{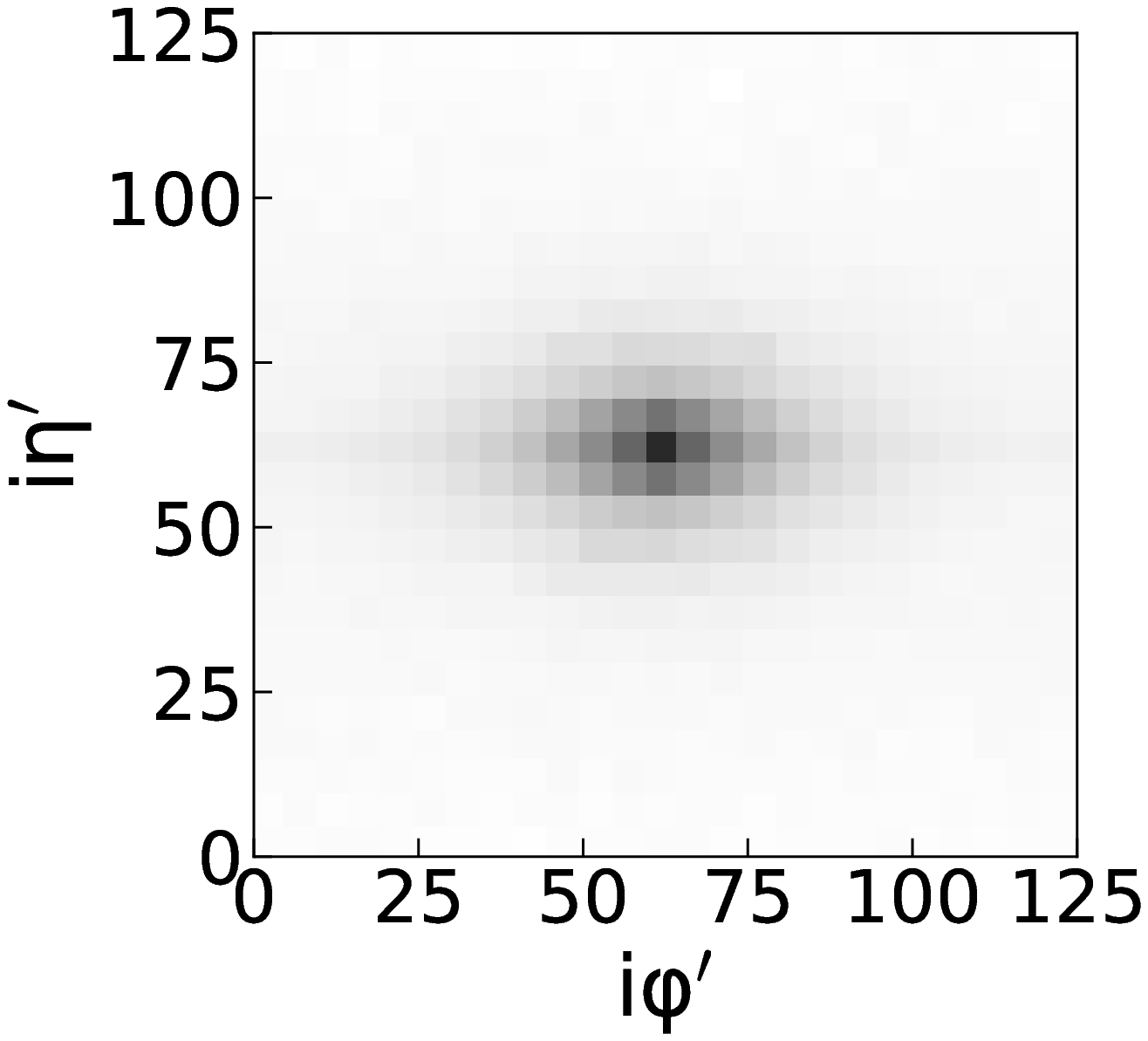}
  \caption{HCAL overlay. Left: gluon jet, Right: quark jet.}
  \label{fig:overlay_2}
\end{subfigure}
\caption{Jet-view image overlays split by subdetector: tracks (\ref{fig:overlay_0}), ECAL (\ref{fig:overlay_1}), and HCAL (\ref{fig:overlay_2}) over 70k jets each, all in log-scale. Gluon jets appear on the left, quark jets on the right. Gluon jet showers are visibly more dispersed in all channels. Note the presence of horizontal bands in the ECAL (\ref{fig:overlay_1}) overlays, highlighting the image fidelity to capture the energy leakage across the ECAL barrel-endcap boundary. Image resolution: $125 \times 125$.}
\label{fig:overlays}
\end{figure}

\begin{figure}[!htbp]
\centering
\begin{subfigure}{.6\textwidth}
  \centering
  \includegraphics[width=.45\linewidth]{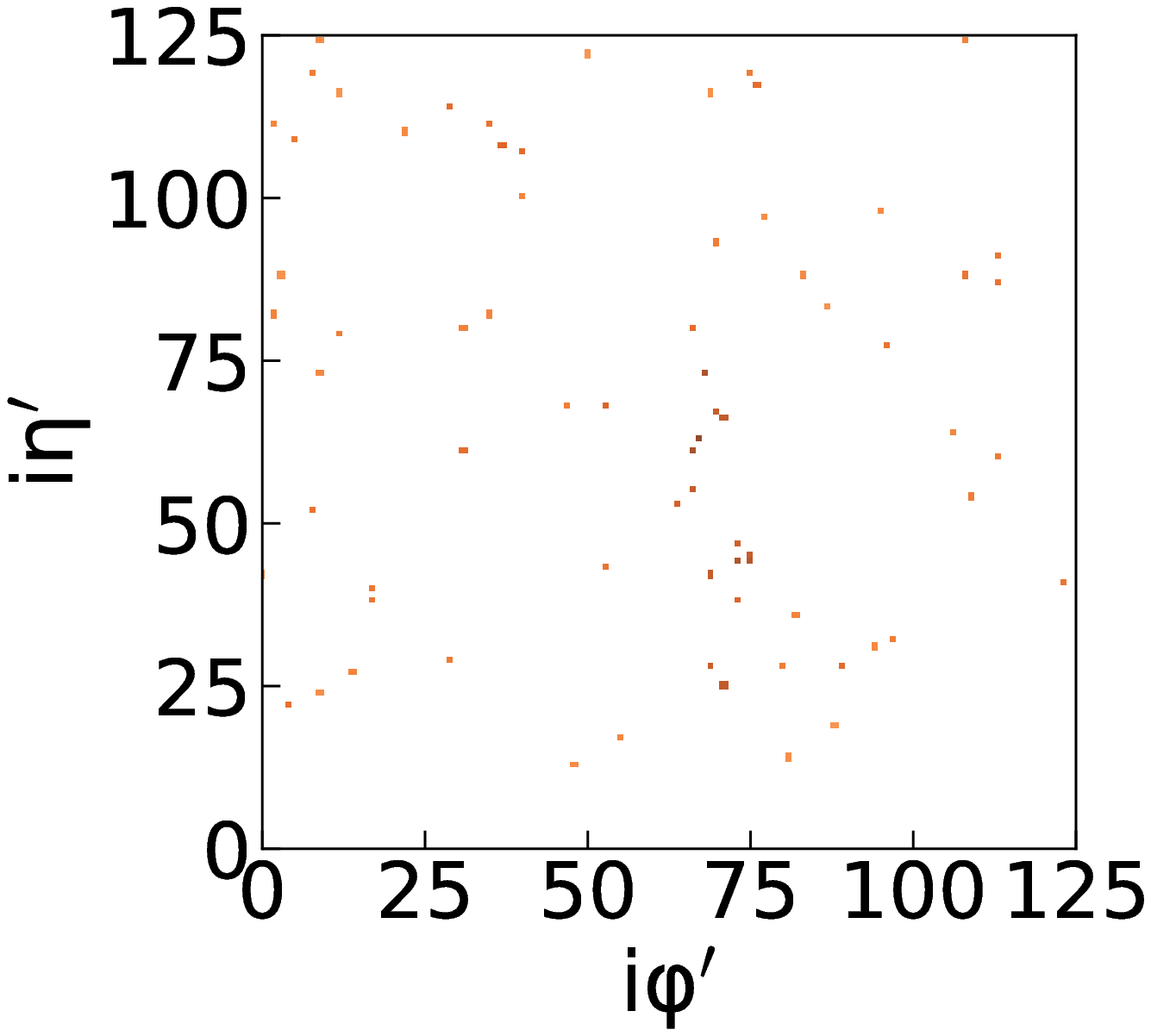}
  \includegraphics[width=.45\linewidth]{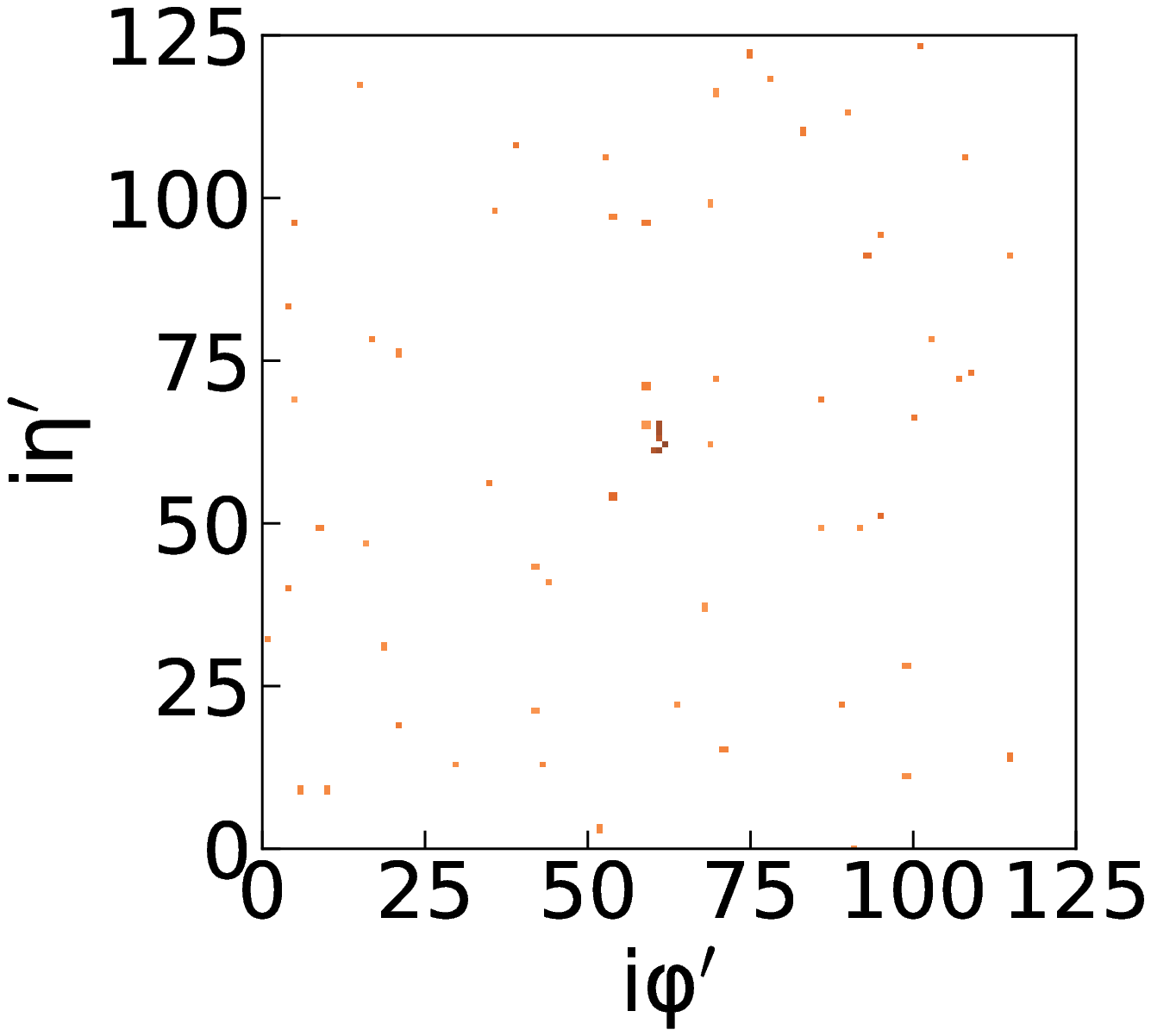}
  \caption{Tracks channel. Left: gluon jet, Right: quark jet.}
  \label{fig:single_0}
\end{subfigure}
\begin{subfigure}{.6\textwidth}
  \centering
  \includegraphics[width=.45\linewidth]{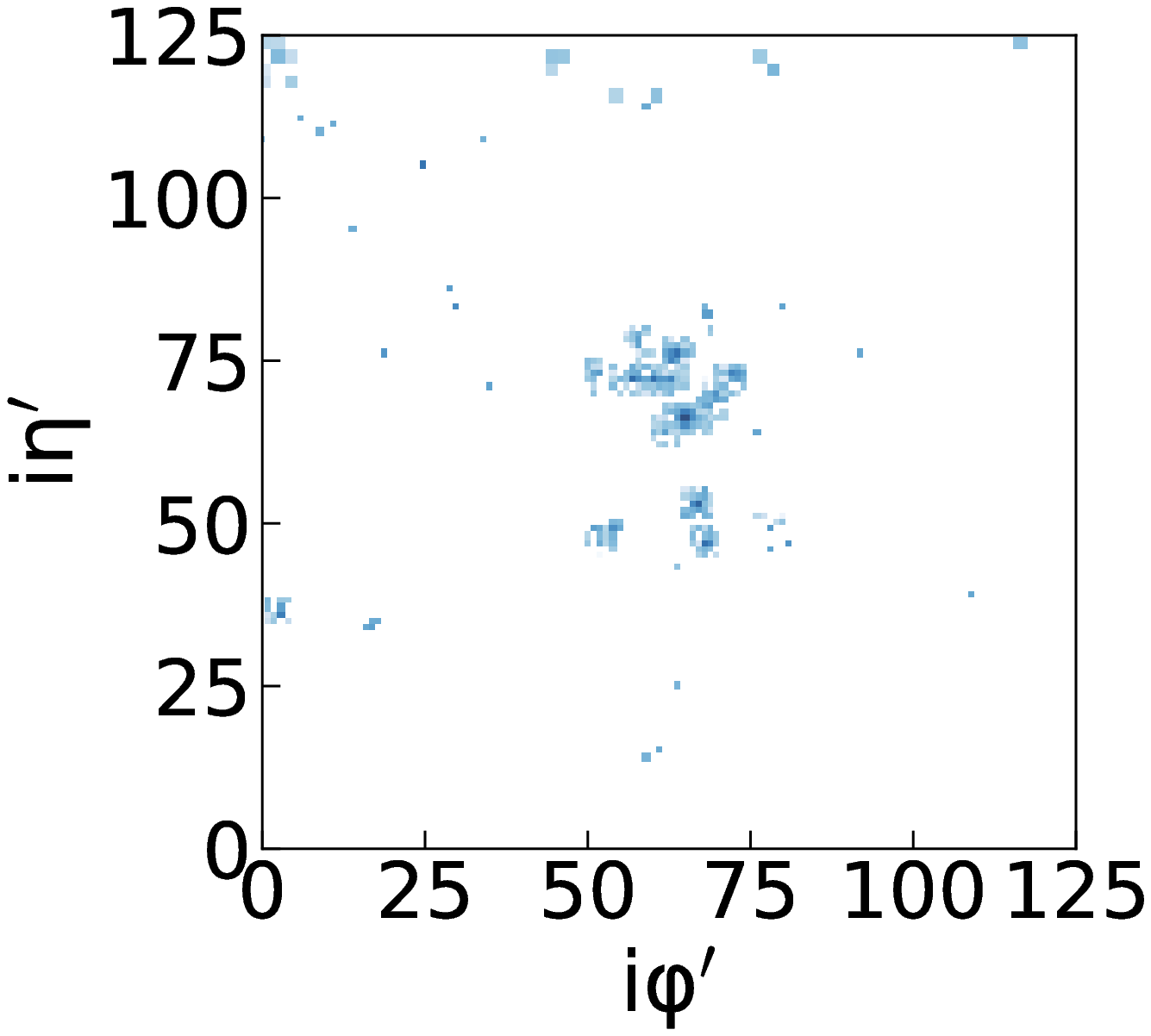}
  \includegraphics[width=.45\linewidth]{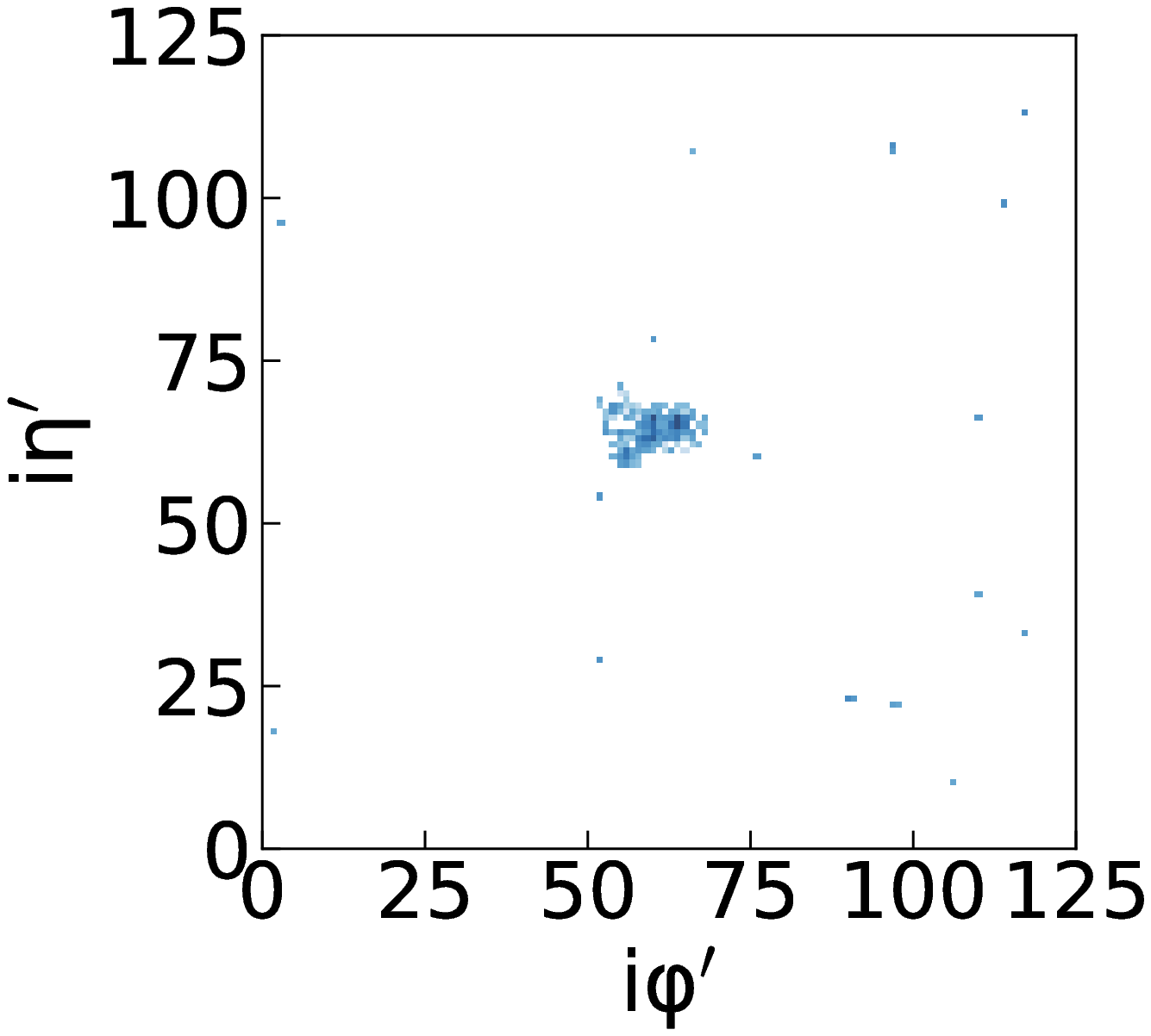}
  \caption{ECAL channel. Left: gluon jet, Right: quark jet.}
  \label{fig:single_1}
\end{subfigure}
\begin{subfigure}{.6\textwidth}
  \centering
  \includegraphics[width=.45\linewidth]{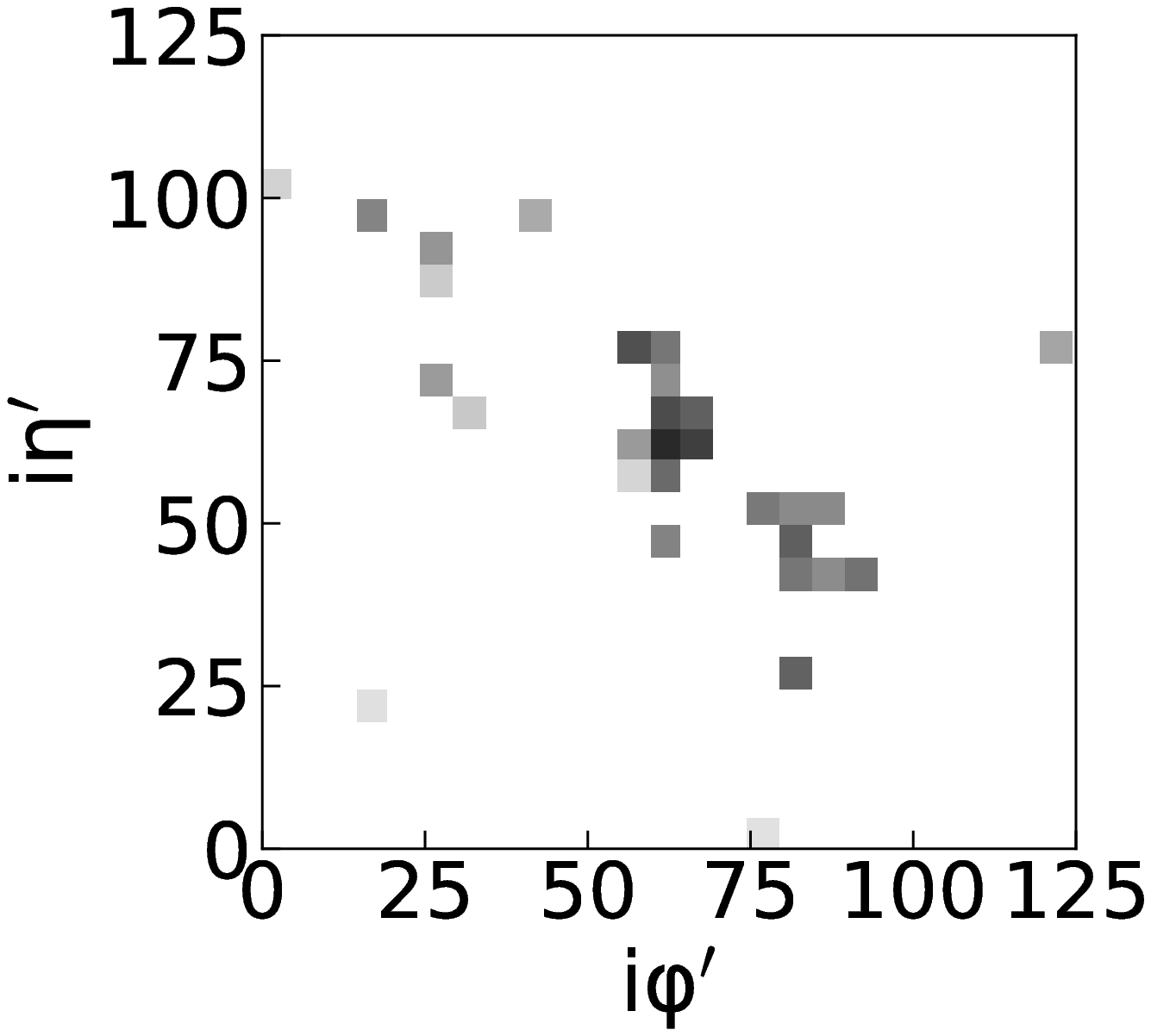}
  \includegraphics[width=.45\linewidth]{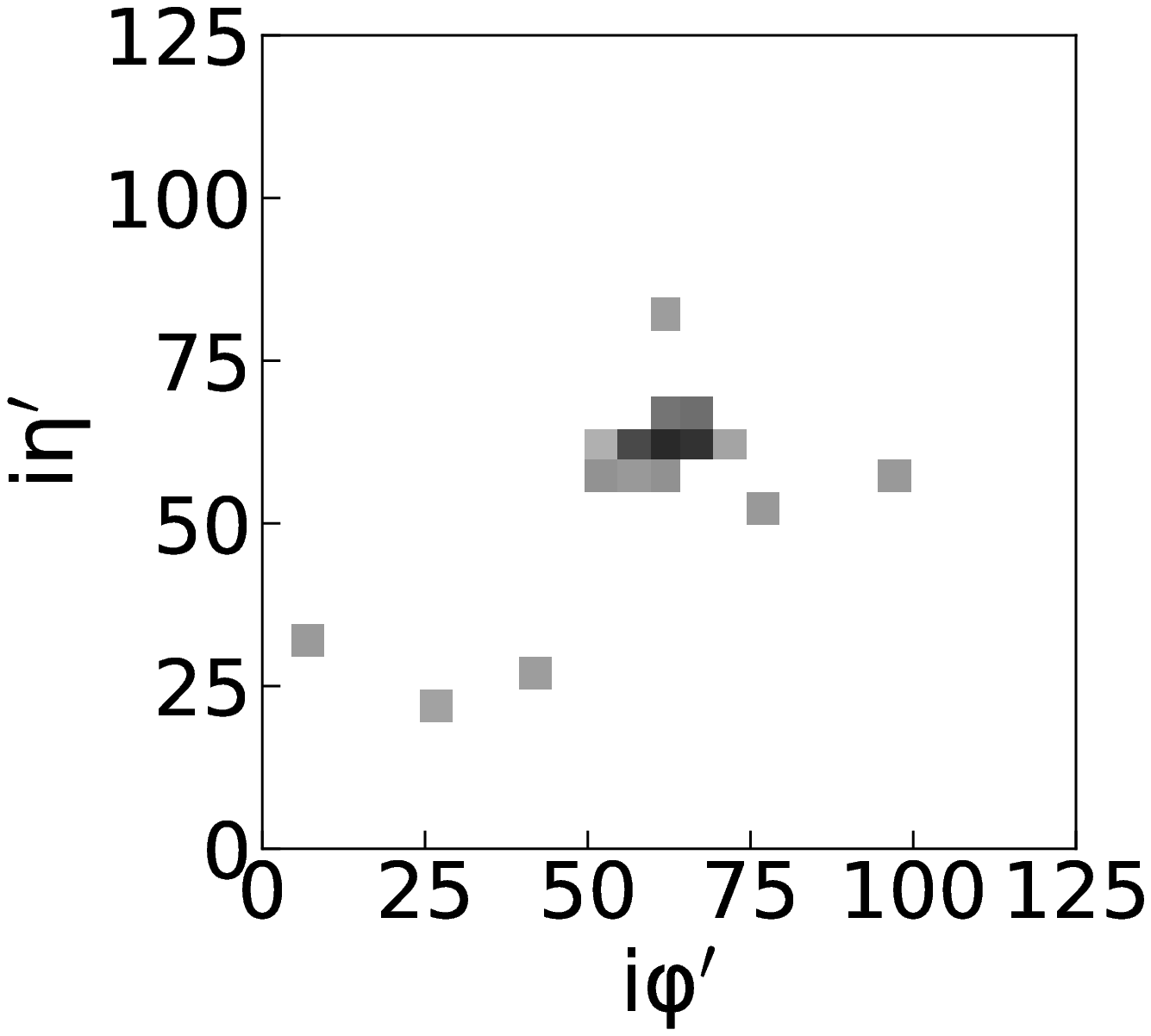}
  \caption{HCAL channel. Left: gluon jet, Right: quark jet.}
  \label{fig:single_2}
\end{subfigure}
\begin{subfigure}{.6\textwidth}
  \centering
  \includegraphics[width=.45\linewidth]{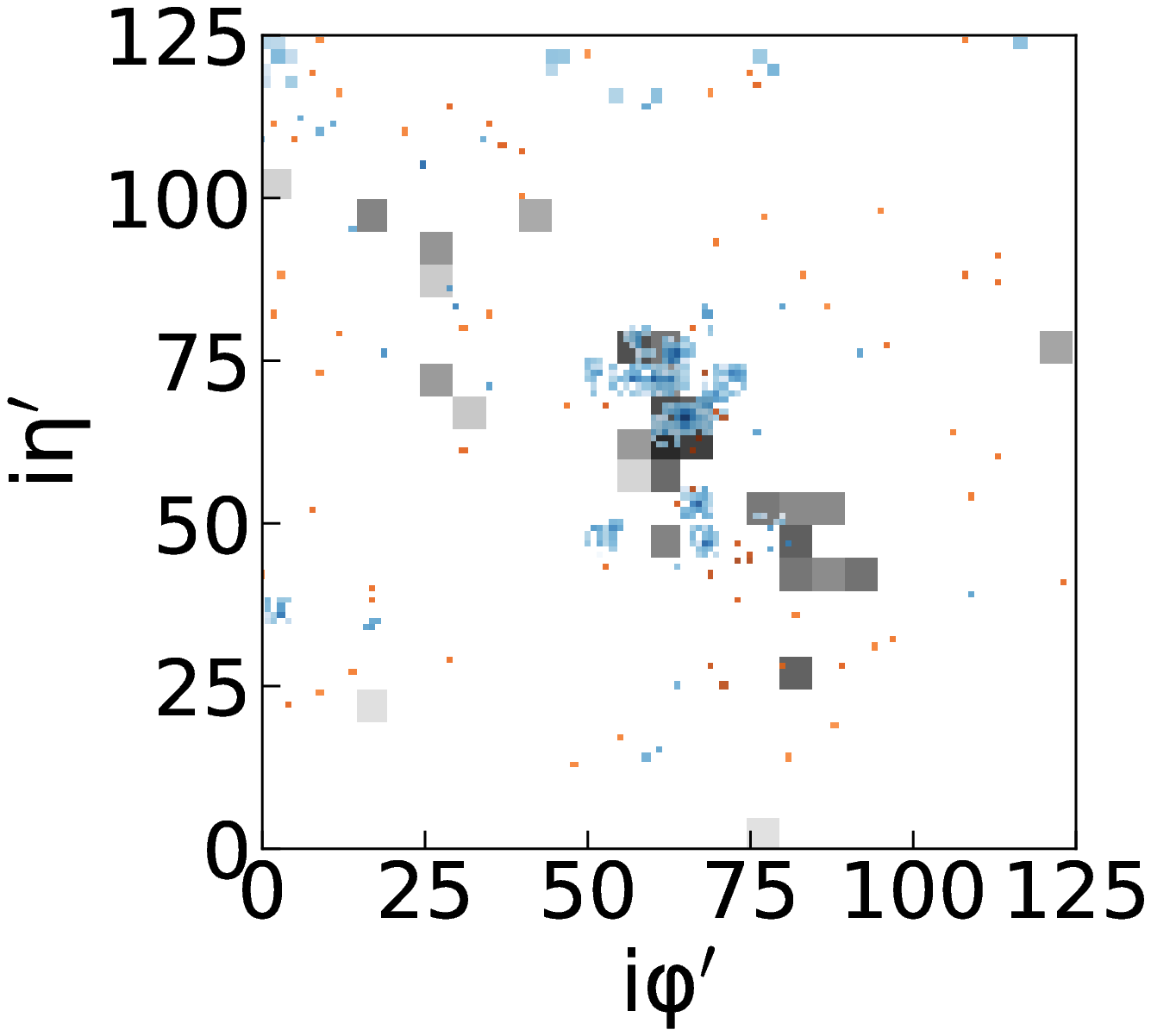}
  \includegraphics[width=.45\linewidth]{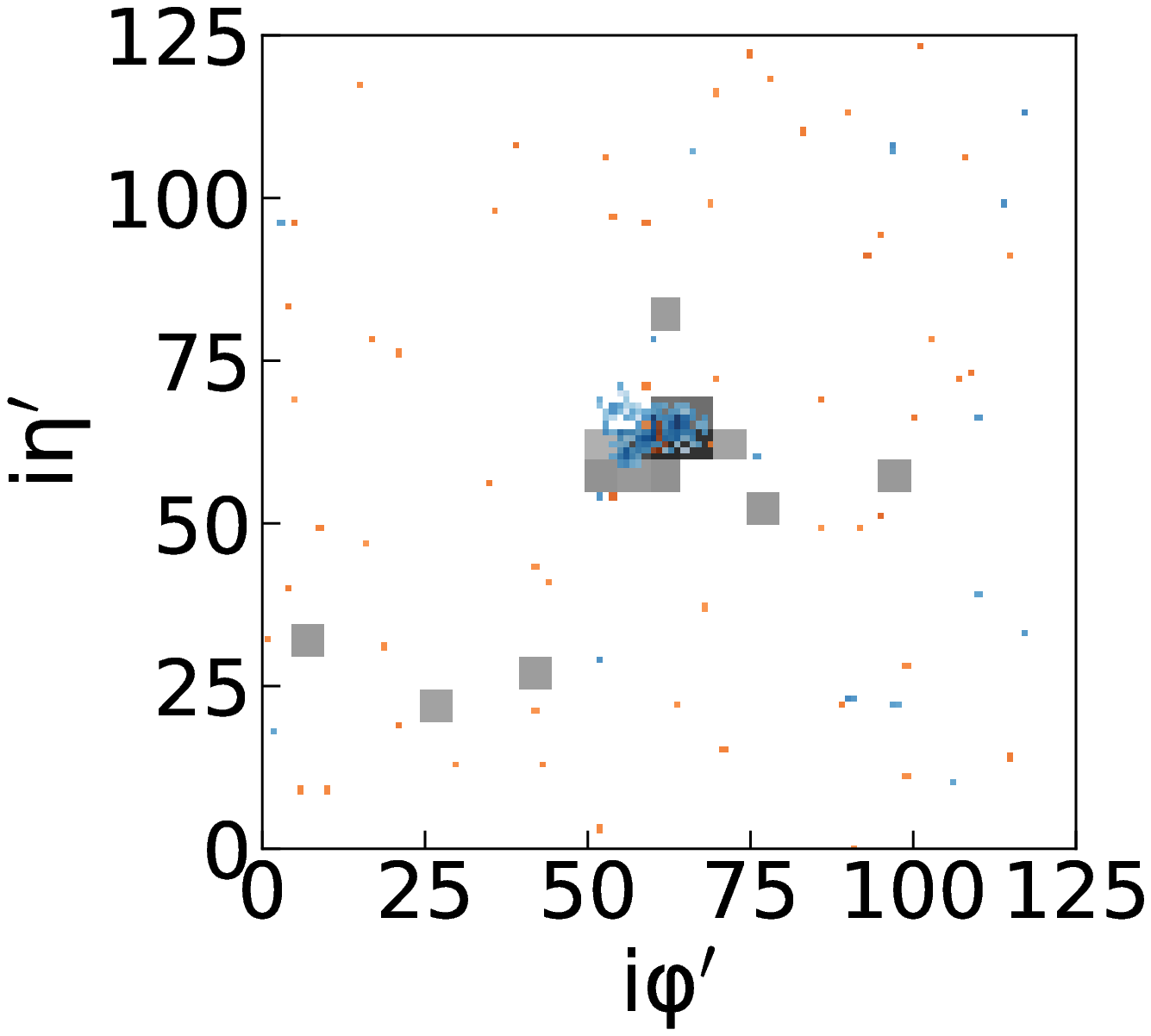}
  \caption{Composite jet image. Left: gluon jet, Right: quark jet.}
  \label{fig:single_3}
\end{subfigure}
\caption{Representative jet-view image for a single jet split by sub-detector: tracks (\ref{fig:single_0}), ECAL (\ref{fig:single_1}), and HCAL (\ref{fig:single_2}) and combined into a composite image (\ref{fig:single_3}), all in log-scale. Gluon jets appear on the left, quark jets on the right. Gluon jet showers are visibly more dispersed in all channels. Image resolution: 125 $\times$ 125.}
\label{fig:singles}
\end{figure}

To gain a better intuition for the jet images, we present a number of visualizations: Figure~\ref{fig:overlays} shows the various subdetector image overlays averaged over the full test set of about 70k jets for each class (described in Table~\ref{table:Nevents}), while Figure~\ref{fig:singles} shows subdetector images for a single jet. Visually, one notes two main differences from previous, particle-based jet images (e.g.~\cite{qvgimg, atlasjetimg}). First, the end-to-end images appear noticeably more ``raw'' in that they contain more noise and stray hits. This is, of course, expected given that we are looking at the (simulated) physical detector deposits in all their richness. Importantly, this provides a prime setting for the classifier to learn an internal noise-mitigation strategy, potentially allowing it to maintain performance even under higher PU conditions. This PU robustness would be greatly diminished if one were to instead train on highly pruned jet images, which could themselves be stripped of meaningful hits. Second, the end-to-end jet images are rendered in the finer ECAL-like granularity as opposed to the coarser HCAL-like granularity more commonly used in previous work. As the results in Section~\ref{sec:jetID} show, quark versus gluon discrimination is essentially dominated by precise spatial resolution. Moreover, while each subdetector channel has the same $125\times125$ resolution, the effective size of the particle features differs dramatically across the subdetector images due to differences in the underlying detector resolution. That is, in the tracks image, particles appear as individual, isolated pixels, while in the ECAL image, as roughly $3\times3$ pixel showers, and in the HCAL image, as clusters of $5\times5$ pixel blocks. Such a classification task, therefore, poses a unique, multi-scale feature extraction task for the CNN. As we will find, CNNs do not disappoint.

As a reference to establish the maximum performance that can be obtained with the ECAL-like granularity detector images in the limit that particle position resolution saturates image resolution, we construct detector images filled with only isolated pixels (i.e. with no lateral shower width) corresponding to the generator-level particle positions. Specifically, we take all the stable particles from the \texttt{Pythia} generator's particle table and construct a two-channel image with pixels corresponding to the ($\eta$,$\phi$)-positions of the stable particles weighted by their $p_T$. These include particles from both the hard-scatter and the underlying event but not from PU. We place all electrons and photons in one image channel and all the hadrons in the other. We then train on jet-view and full detector-view images as before.

\section{Network and Training\label{sec:Net}}

We use the same training strategy both for the classification of quark versus gluon jets, and for the classification of di-quark versus di-gluon QCD events. We employ the ResNet-15~\cite{resnet} CNN architecture found in~\cite{e2e}, where the same hyper-parameters were found to be optimal. The ADAM adaptive learning rate optimizer~\cite{adam} is used to minimize the binary cross-entropy loss in batches of 32 samples. We use an initial learning rate of $5\times 10^{-4}$, and explicitly reduce it by half every 10 epochs for a total of 30 training epochs. We reserve about 26k out of the 768k training samples for our validation set (see Table~\ref{table:TrainSplit}). All training was done using \texttt{PyTorch}~\cite{pytorch} running on a single NVIDIA Titan X GPU with \texttt{PyArrow} for data I/O~\cite{pyarrow}.

\begin{table}[h]
\centering
\begin{tabular}{c c c}
\hline
\textbf{Training Samples} & \textbf{Validation Samples} & \textbf{Test Samples} \\
per class        		  & per class          & per class \\
\hline
384000                    & 12950              & 69653\\
\hline
\end{tabular}
\caption{Number of events in training, validation, and final test set \emph{for each class}. The total training+validation and test sets contain a balanced proportion of class samples.}
\label{table:TrainSplit}
\end{table}

\begin{table}[h]
\centering
\begin{tabular}{l l l}
\hline
\textbf{Algorithm} & \textbf{Inputs}        & \textbf{Architecture} \\
\hline
A                & 2 $\times$ $q/g$ score     & FCN128$\times$2 \\
B                & 2 $\times$ $q/g$ score, jet 4-momenta & FCN128$\times$2 \\
C                & Full detector image   & ResNet-15 \\
\hline
\end{tabular}
\caption{List of end-to-end event classification algorithms.}
\label{table:evt_nets}
\end{table}
For the jet classification studies, we use the multi-subdetecor channel jet-view images as constructed in Section~\ref{sec:Images} and feed these directly to the CNN. To aid in the interpretation of the results, we also experiment with different combinations of subdetector channels in the input image, as discussed in greater detail in Section~\ref{sec:jetID}.

For the event classification studies, we explore a number of approaches for integrating jet classification into a dijet event classification workflow, as summarized in Table~\ref{table:evt_nets}. Naively, the simplest way to construct a dijet classifier would be to simply concatenate the outputs of the individual jet classifiers, ordered by $p_T$, and feed these into a fully-connected neural network (FCN) which acts as the actual event classifier. The jet classifier output used as input to the event classifier could
either be the final feature vector of the CNN, or just the final quark versus gluon prediction ($q/g$ score). While the latter tends to be more standard among LHC analyses, the former could potentially be beneficial if the event discrimination were driven by some other internal property of the jet unrelated to its $q/g$ score. In this case, it is possible to either train the jet classifier in-place concurrently with the event classifier, or use a pre-trained, pre-calibrated jet classifier. Since such a scenario is not expected for dijet classification, we simply use the $q/g$ score from the best end-to-end jet classification algorithm as input to a FCN-based dijet event classifier (algorithm A). This, however, does not account for the event-level kinematics of the jet. Therefore, we can additionally augment the dijet classifier inputs with the 4-momenta of the reconstructed jets (or the coordinates of the jet image centers), also ordered by $p_T$ (algorithm B). 

The above event classifiers collectively follow a factorized workflow, where the jet identification is done first, followed by a separate event classifier that is engineered to exploit the topology of the underlying physics process. However, one could also construct a unified end-to-end event classifier that takes in the full detector-view image (see Figure~\ref{fig:fullye2e}) all at once (algorithm C) to perform dijet classification directly from detector date. This avoids the particle ordering and numbering issues introduced in Section~\ref{sec:Introduction} since the natural spatial distribution of the jets in the full detector is exploited. We found no significant benefit to augmenting the full detector-view dataset with images randomly ``rotated'' in $\phi$ to enforce azimuthal symmetry, suggesting this had been learned by the classifier implicitly.

Algorithm C may be particularly suited to complex, multi-body decays where engineering an effective event classifier is non-trivial. Indeed, because algorithm C does not depend on one's ability to model the event topology, it can potentially serve as a guide for engineering a specialized one that does. One challenge in implementing algorithm C in an analysis is calibrating and deriving uncertainties for it, since jet- and event-level sources of uncertainty become coupled into a single classifier response. A possible solution is to first derive the jet-level sources by selectively masking out the detector image outside the object-of-interest. These studies can then be used to correct the response of the classifier before deriving the event-level sources on the fully unmasked detector image. More detailed, analysis-specific work is needed to demonstrate this approach.

To evaluate classifier performance, we use the Receiver Operating Characteristic (ROC) curve, which can be interpreted in terms of the signal efficiency (true positive rate) versus background rejection (true negative rate), as is commonly used in high-energy physics. The area under the ROC curve (AUC) is used to select the best algorithm based on the validation set. In addition, we also present the inverse of the false positive rate (FPR) at a fixed true positive rate (TPR) of 70\%. For an unbiased estimate of performance, all final performance metrics presented in our Results (Sections~\ref{sec:jetID},\ref{sec:eventID}) are calculated from the test set which is statistically independent from the validation set.

\section{Jet ID Results\label{sec:jetID}}

The end-to-end jet classification results lend themselves well to a detector performance interpretation. To this end, we first perform the quark versus gluon tagging using single subdetector-channel images, to understand the relative importance of each subdetector. The results are presented in the last three rows of Table~\ref{table:JetIDresults}. The best single subdetector performance comes from the reconstructed tracks image followed by ECAL, then HCAL, correlating strongly with the resolution of the underlying physical detector. Given that quark versus gluon jets differ primarily in how broadly their jet constituents are spaced, increased detector resolution would mean we are better able to resolve the positions of the constituents. In fact, a close visual inspection of the core of the single-jet subdetector images in Figure~\ref{fig:singles} suggests this pattern of broader constituent distribution is indeed more apparent with the finer tracks image, followed by the ECAL image where the constituents are just barely resolved for the gluon jet but not for the quark jet. By the HCAL image, it is no longer so obvious. This reinforces the importance of building high-fidelity, full-granularity detector images that are able to capture all the nuances in the energy deposition patterns. Lastly, we note the remarkable ability of the CNNs to extract meaningful information even from the highly-sparse tracks images which is purely composed of isolated pixels if not empty space.

\begin{table}[h]
\centering
\begin{tabular}{l c c}
\hline
& & \textbf{1/FPR} \\
\textbf{Jet image} & \textbf{ROC AUC} & \textbf{@TPR=0.7} \\
\hline
Generated EM+Had & 0.854 & 6.46 \\
\hline
Tracks+ECAL+HCAL & 0.807 & 4.45 \\
\hline
Tracks+ECAL      & 0.804 & 4.35 \\
ECAL+HCAL        & 0.781 & 3.76 \\
\hline
Tracks           & 0.782 & 3.73 \\
ECAL             & 0.760 & 3.28 \\
HCAL             & 0.682 & 2.27 \\
\hline
\end{tabular}
\caption{End-to-end jet classification results. Statistical uncertainties are in the third significant figure.}
\label{table:JetIDresults}
\end{table}

We next consider the effect of combining two subdetector images in a single multi-channel image, as presented in rows 3-4 of Table~\ref{table:JetIDresults}. We can combine the tracks and ECAL images to incorporate information about the photons that are absent from the tracks image (Tracks+ECAL). Alternatively, we could swap out the tracks image for the HCAL (ECAL+HCAL), which amounts to taking the charged hadron information from the coarser HCAL image, to form a purely calorimetric image. The former approach achieves the best discrimination so far, while the latter is less performant due to having sacrificed the precise spatial information of the tracks. Indeed, despite the ECAL+HCAL image having the advantage in terms of neutral hadron information, we see that the tracks-only image performs as well as the calorimeter-only image, highlighting again the importance of precise spatial resolution.

Of course, the best overall performance is obtained when all three subdetector images are combined together into a single input image (Tracks+ECAL+HCAL), as shown in the second row of Table~\ref{table:JetIDresults}. Although the three subdetector images have identical image resolution, the effective feature scales among them differ dramatically (see Figure~\ref{fig:singles}). Therefore, that the CNN can extract meaningful features at these different feature scales and deliver robust performance notwithstanding serves as a testament to their power and versatility.

We can then compare these results to the generator-level images to understand how much better a jet classifier could perform in the limit of maximal detector resolution and no PU. From the top row of Table~\ref{table:JetIDresults} (Generated EM+Had), we find that detector resolution effects and PU together account for about 5\% of the performance loss at the jet level.

\begin{figure}[!htbp]
\centering
\includegraphics[width=.45\linewidth]{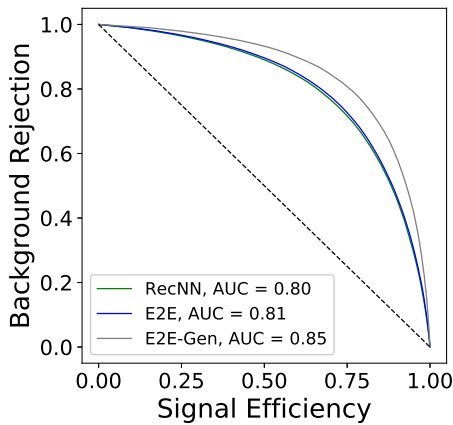}
\caption{Jet classification ROC curves.}
\label{fig:jetIDrocs}
\end{figure}

To put the end-to-end jet classification results into context, we can benchmark against one of the current state-of-the-art jet classifiers, the QCD-aware recursive neural network (RecNN) algorithm~\cite{kyle, taoli}. We use the default architecture, hyper-parameters, and training strategy implemented in~\cite{recnngithub} but with the training split and evaluation frequency modified for consistency with the strategy used here (see Section \ref{sec:Net}). Using the exact same jet samples as used for the end-to-end studies, we find the following results between the different pre-processing schemes in Table~\ref{table:JetIDbenchmark}. For the top-scoring pre-processing scheme, we calculate the mean and standard deviation over 5 random seeds of the training set shuffling (as in~\cite{kyle}), and compare this with the Tracks+ECAL+HCAL jet image results. These appear in the top two rows of Table~\ref{table:JetIDbenchmark} with the corresponding ROC curves plotted in Figure~\ref{fig:jetIDrocs}.

\begin{table}[h]
\centering
\begin{tabular}{l c c}
\hline
& & \textbf{1/FPR} \\
\textbf{Jet ID Algorithm} & \textbf{ROC AUC} & \textbf{@TPR=0.7}\\
\hline
Jet image, Tracks+ECAL+HCAL  & 0.8076 $\pm$ 0.0002  & 4.473 $\pm$ 0.013\\
\hline
RecNN, ascending-$p_T$           & 0.8017 $\pm$ 0.0002  & 4.281 $\pm$ 0.013\\
RecNN, descending-$p_T$          & 0.802                & 4.30\\
RecNN, anti-$k_T$                & 0.800                & 4.25\\
RecNN, Cambridge/Aachen          & 0.801                & 4.26\\
RecNN, no rotation/re-clustering & 0.800                & 4.23\\
RecNN, $k_T$                     & 0.800                & 4.24\\
RecNN, $k_T$-colinear10-max      & 0.799                & 4.23\\
RecNN, random                    & 0.797                & 4.15\\
\hline
\end{tabular}
\caption{End-to-end versus RecNN jet classification results. Top scores for each algorithm represent mean and standard deviation over 5 trials. Statistical uncertainties are in the third significant figure.}
\label{table:JetIDbenchmark}
\end{table}

We find the end-to-end jet image algorithm to be highly competitive with the top performing RecNN, even after taking into account statistical uncertainties and variations due to the random number seed. While the ascending-$p_T$ pre-processing gives the best RecNN results, most of the other schemes fare comparably. As previous studies~\cite{deepjet, kyle} have shown image-based approaches to underperform relative to direct particle-data based algorithms, our results suggest this discrepancy can be attributed to limitations in the jet image construction rather than to the use of CNNs themselves. Since previous jet image work involved pixelating the particle-level jet constituents into HCAL-like granularity images, this is not unexpected. High-fidelity, high-granularity detector images that are as minimally processed and as information rich as possible, therefore, are essential to effective jet image tagging. For heavy jet flavor tagging, in particular, further progress is likely to be found in more sophisticated treatments of the tracking detectors.

\section{Event ID Results\label{sec:eventID}}

\begin{figure}[!htbp]
\centering
\includegraphics[width=.45\linewidth]{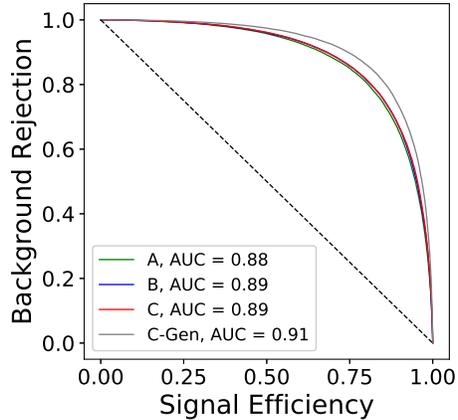}
\caption{Event classification ROC curves.}
\label{fig:eventIDrocs}
\end{figure}

We can extend quark versus gluon tagging to QCD di-quark versus di-gluon event classification to compare the different ways one might build a dijet event classifier using end-to-end jet classification, as described in Section~\ref{sec:Net}. The results are summarized in Table~\ref{table:E2EEventIDresults} with the corresponding ROC curves in Figure~\ref{fig:eventIDrocs}. The result of using generator-level particle images (algorithm C-Gen) is also included for reference.

\begin{table}[h]
\centering
\begin{tabular}{l c c}
\hline
& & \textbf{1/FPR} \\
\textbf{Event ID Algorithm} & \textbf{ROC AUC} & \textbf{@TPR=0.7}\\
\hline
\textbf{A:} 2 $\times$ $q/g$ score
    & 0.882 & 8.44\\
\textbf{B:} 2 $\times$ $q/g$ score $+$ jet 4-momenta
    & 0.888 & 9.04\\
\textbf{C:} Full detector image
    & 0.889 & 9.05\\
\textbf{C-Gen:} Full detector image,\\
    generated particle inputs & 0.911 & 12.92\\
\hline
\end{tabular}
\caption{End-to-end event classification results. Statistical uncertainties are in the third significant figure.}
\label{table:E2EEventIDresults}
\end{table}

These indicate that event classification performance is dominated by jet-level differences (algorithm A), with negligible gain from including the jet 4-momenta (algorithm B). To first approximation, this is expected given that both di-quark and di-gluon production channels have similar non-resonant kinematics. Although di-quark events are slightly differentiated in their angular distribution due to spin correlation and polarization effects, these effects are not expected to be of significance. The results for algorithm B were not sensitive to the choice of using the reconstructed jet centroids or the image coordinates of the jet image centers. While the full detector-view approach (algorithm C) performs just as well as the topology-specific approach of B, this confirms C's ability to learn the same information as B at both the jet- and event-level, without prior knowledge of the event topology. The approach of C, therefore, may prove illuminating when tackling more complex decays which are difficult to model. Lastly, looking at the results of the generator-level images with no PU, we find the difference relative to the physical detector images to be approximately 3\% at the event level versus 5\% at the jet level, likely owing to the complementary information that two jets provide.

To gain a better understanding of the impact of the underlying event and PU in the full detector-view approach (algorithm C), we train a separate event classifier on full detector images but with the pixel intensities outside of the jet windows masked or zeroed out (algorithm C-Zero). We can then perform a transfer learning test by evaluating the classifier trained on the original scenario (algorithm C) on the zeroed-out detector images and vice-versa. The results are presented in Table~\ref{table:E2EEventIDzero}.

\begin{table}[h]
\centering
\begin{tabular}{l c c}
\hline
 & & \textbf{1/FPR} \\
\textbf{Event ID Algorithm} & \textbf{ROC AUC} & \textbf{@TPR=0.7}\\
\hline
\textbf{C:} Full detector image 
                 & 0.889 & 9.05\\
\textbf{C-Zero:} Full detector image, \\
jet windows only & 0.887 & 8.95\\
\textbf{C}, evaluated on \textbf{C-Zero}
                 & 0.883 & 8.56\\
\textbf{C-Zero}, evaluated on \textbf{C}
                 & 0.884 & 8.58\\
\hline
\end{tabular}
\caption{Event classification supplementary results. Statistical uncertainties are in the third significant figure.}
\label{table:E2EEventIDzero}
\end{table}

As these indicate, the loss in performance from either training starting point is minimal, showing that the end-to-end algorithm is largely insensitive to the underlying event and pileup outside of the jet region-of-interest, suggesting the classifier is focusing solely on the pertinent features of the event. This suggests the end-to-end technique is effective at PU mitigation and detector images should be as minimally processed as possible when presented to the CNN in order to maximize the information that can be extracted.

\section{\label{sec:Conclusions}Conclusions}

In this paper, we demonstrated the application of end-to-end classification techniques to the tagging of light quark jets versus gluon jets using simulated CMS Open Data. We constructed jet-view images, which were high-fidelity maps of the physical detector deposits, to give the jet classifiers direct access to the maximum recorded event information. The resulting multi-subdetector images showed rich features spanning various length scales. Using a ResNet-15 CNN, we achieved effective feature extraction to obtain performance competitive with current state-of-the-art quark versus gluon taggers based on traditional particle-level inputs. We found that precise spatial resolution was of paramount importance, highlighting the importance of high-fidelity detector images and especially the critical role played by the track information. We also explored classifying di-quark versus di-gluon QCD events to illustrate ways in which end-to-end jet classifiers can be used to build event classifiers. We found the discriminating performance to be largely dominated by jet-level differences and thus noted similar performance across the different dijet classifier architectures. 

Finally, we showed that full detector-view event classifiers were robust and versatile against underlying event and pileup outside the jet region-of-interest, making them a compelling tool for complex, multi-body event topologies, where event classifier engineering becomes non-trivial. In future work, we plan to pursue more sophisticated representations of the tracking information in the context of heavy flavor jet tagging and boosted decays.

\section*{\label{sec:Acknowledgments}Acknowledgments}
We thank the entire CMS Collaboration for successfully recording LHC
proton-proton collision data as well as producing and releasing high
quality simulated data used in this paper. We also congratulate all
members in the CERN accelerator departments for the excellent
performance of the LHC and thank the technical and administrative staffs
at CERN and at other CMS institutes for their contributions to the
success of the CMS effort. In addition, we gratefully acknowledge the
computing centres and personnel of the Worldwide LHC Computing Grid for
delivering so effectively the computing infrastructure essential to CMS
analyses. Finally, we acknowledge the enduring support for the
construction and operation of the LHC and the CMS detector.

We would like to thank the CMS Collaboration and the CERN Open Data group for releasing their simulated data under an open access policy. We strongly support initiatives to provide such high-quality simulated datasets that can encourage the development of novel but also realistic algorithms, especially in the area of machine learning. We believe their continued availability will be of great benefit to the high energy physics community in the long run.

Finally, M.A.~and M.P.~are supported by the Office of High Energy Physics of the U.S.~Department of Energy (DOE) under award DE-SC0010118. S.A.~is supported by the Marie Sk\l odowska-Curie Innovative Training Network Fellowship of the European Commission’s Horizon 2020 Programme under contract number 765710 INSIGHTS.


\end{document}